\documentclass[journal=jacsat,manuscript=article]{achemso}
\usepackage[version=3]{mhchem}
\usepackage{booktabs}
\usepackage{threeparttable}
\usepackage{multirow}
\usepackage{siunitx}
\usepackage{amssymb}
\usepackage{amsfonts}
\usepackage{rotating}
\usepackage{array}
\usepackage{amsmath}
\usepackage{bm}

\author{Pawel Dabrowski-Tumanski}
\affiliation[WUT]{Faculty of Mathematics and Information Science, Warsaw University of Technology, Warsaw 00-662, Poland}
\alsoaffiliation[UKSW]{Faculty of Mathematics and Natural Sciences, University of Cardinal Stefan Wyszynski, Warsaw 01-938, Poland}
\alsoaffiliation[Ingenix]{Ingenix.AI}
\email{pawel.tumanski@pw.edu.pl}

\author{Bartosz Topolski}
\affiliation[Ingenix]{Ingenix.AI}

\author{Dariusz Plewczynski}
\affiliation[WUT]{Faculty of Mathematics and Information Science, Warsaw University of Technology, Warsaw 00-662, Poland}
\alsoaffiliation[UW]{Centre of New Technologies, University of Warsaw, Warsaw 02-097, Poland}
\alsoaffiliation[Ingenix]{Ingenix.AI}

\author{Tomasz Jetka}
\affiliation[Ingenix]{Ingenix.AI}

\newcolumntype{L}[1]{>{\raggedright\arraybackslash}p{#1}}

\title[]{The Geometry of Activity Cliffs: Representation Dependence and Multi-Scale Characterization of Activity Landscapes}

\begin{document}

\begin{abstract}
Activity cliffs, structurally similar compounds with large potency differences, are widely treated as intrinsic features of chemical datasets. 
We argue that apart from target biology, much of our cliff understanding is a consequence of the geometry induced by the chosen molecular representation, not a property of a molecule pair itself. 

We designed a six-step pipeline to systematically test this hypothesis. The pipeline consists of: assessing pairwise distance geometry, cliff enrichment, activity gradient distribution, persistent homology of the cliff subspace, predictive benchmarking for a chosen pair of an embedding and a metric, and eventually, analysis of the matched molecular pairs and stereoisomers.
We applied the pipeline to fifteen configurations of embeddings and metrics to build a benchmark across three distinctive datasets known of activity cliffs challenges.

No representation excels on all criteria: Morgan Tanimoto provides the strongest cliff enrichment and cross-scaffold generalization; MolFormer cosine provides the only meaningful stereochemical sensitivity; MACCS and RDKit Dice fingerprints are most sensitive to matched-molecular-pair transformations; ChemBERTa fails uniformly due to embedding collapse. 

These findings are not a ranking. They reflect the fact that different representations encode different aspects of molecular recognition, and that choosing one implicitly defines what an activity cliff actually is.
\end{abstract}


\section{Introduction}

Activity cliffs are pairs of structurally similar compounds with large differences in biological potency that 
demarcate the boundaries of predictability in structure-activity relationships.
\cite{maggiora2006outliers,stumpfe2012exploring,stumpfe2014recent} Although the concept is now ubiquitous 
in medicinal chemistry and cheminformatics, its definition is operational rather than formal: it depends on 
two user-chosen thresholds: a potency gap typically of one log unit and a similarity cutoff. The latter
requires an implicit choice of how molecular similarity is measured at all. This rests on an 
unexamined foundational choice: the selection of a molecular representation and a similarity metric.

Indeed, while the target biology dictates the baseline roughness of the activity landscape, it is metric which 
dictates the arbitrary boundaries of the cliff, as the structural proximity is not an intrinsic property of a 
molecule pair, but rather a property of the metric space in which the molecules are embedded. A concrete 
illustration: on a single bioactivity dataset, the number of pairs that qualify as activity cliffs at a fixed 
similarity cutoff can vary by more than an order of magnitude -- in our experiments by a factor of roughly 
50 -- across reasonable choices of representation, even though the compounds and the potency measurements 
are identical. An even sharper case is stereoisomers. In many commonly used 2D fingerprints, stereoisomeric 
pairs can collapse to identical or near-identical representations despite differing only in three-dimensional 
configuration. If their potencies differ, such pairs expose a representation-dependent discontinuity: they may 
appear as duplicate-label conflicts, extreme activity cliffs, or not cliffs at all. In neither 
case has the chemistry changed --- only the metric in which the chemistry is read. Changing the embedding or 
similarity metric changes which pairs qualify as cliffs, how many exist, and whether they are predictable, 
without changing the underlying biology.

This observation has a precise mathematical formulation. Let $(\mathcal{M}, d)$ be a metric space of chemical 
compounds and $\varphi : \mathcal{M} \to \mathbb{R}$ an activity functional. The local Lipschitz constant 
$L(x,y) = |\varphi(x) - \varphi(y)| / d(x,y)$ measures landscape roughness at the pair level; a pair constitutes 
an activity cliff if and only if $L(x,y) > A/s$, where $A$ and $s$ are the activity and similarity thresholds. 
$L(x,y)$ is the Structure-Activity Landscape Index (SALI) of Guha and Van Drie,\cite{guha2008structure} 
reinterpreted as the empirical Lipschitz constant under $d$. Both the cliff indicator and $L(x,y)$ are explicit 
functions of the representation. 

Despite this foundational role, the representation-dependence has not been systematically studied: the field has 
converged on Morgan/Tanimoto as a near-universal default 
\cite{rogers2010extended,willett1998chemical,maggiora2014molecular}, while learned molecular embeddings have 
developed in parallel, without a principled framework for comparing how different representations organize the 
activity landscape.

\begin{figure}
    \centering
    \includegraphics[width=0.5\linewidth]{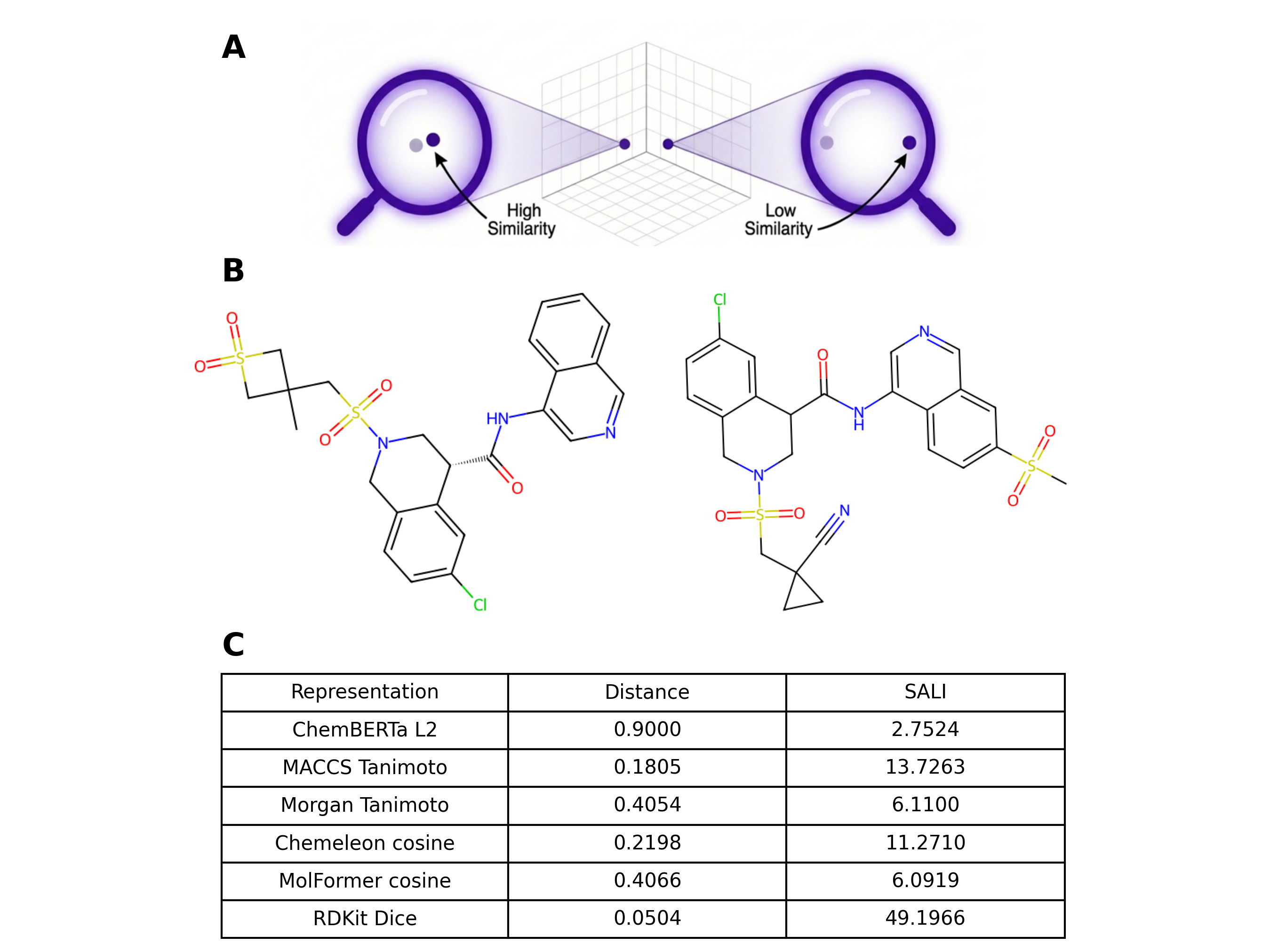}
    \caption{The exemplification of the metric-dependence. (A) The molecule pairs within different representations
    may seem artificially close. (B) An example of a molecule pair from SARS-CoV-2 dataset with activity difference
    $|\Delta\text{pK}_i| = 1.16$. (C) The distances between the molecules in different representations and 
    corresponding SALI index. Depending on the representation, the pair may be classified as an activity cliff, 
    or not a cliff at all.}
    \label{fig:intro}
\end{figure}

The present work provides that framework. We make three contributions. \textbf{Conceptually}, we establish 
that activity cliffs are convolution of representation geometry (embedding and metric) and target 
biology, not just intrinsic properties of molecule pairs. \textbf{Methodologically}, we introduce a six-step 
pipeline in which each step probes a geometrically distinct property of the activity landscape: discriminative 
range, cliff enrichment, gradient tail behavior, topological organization of the cliff subspace, and structure 
of cliff-relevant information in embedding space, followed by real-life validation on matched molecular pairs 
and stereoisomers. In particular, we train different models predicting the existence of activity cliffs.
However, we do not aim in selecting the best possible model, but rather to use the performance gaps as 
diagnostic statistics about the embedding. \textbf{Empirically}, we apply the pipeline to fifteen 
(embedding, metric) configurations spanning classical fingerprints and three learned embeddings across three 
protein targets.

Our thesis is, that selecting a molecule representation without characterizing its geometric properties produces 
conclusions about activity landscapes that may reflect the choice of metric rather than the underlying biology 
(Figure~\ref{fig:intro}). 
Therefore, our goal is not to identify the best representation, but rather to understand how different 
representations encode different and partially non-overlapping aspects of molecular recognition. Especially, as 
it turns out, that no single configuration captures all of them.

\section{Methods}

\subsection{Data}

We used three publicly available bioactivity datasets curated for activity cliff 
research.\cite{dablander2023exploring} The SARS-CoV-2 Main Protease (Mpro) set comprises 1\,924 compounds 
with $K_i$ values from the PostEra Moonshot initiative,\cite{boby2023open} in which we right-censored 
the measurements with $K_i \geq 99\;\mu$M. 
The Factor Xa and Dopamine D2 receptor sets comprise 3\,574 and 6\,280 compounds respectively, 
both drawn from ChEMBL.\cite{mendez2019chembl} All $K_i$ values were converted to 
$\text{pK}_i = -\log_{10}(K_i)$. Pairs with two censored 
measurements were excluded; mixed pairs were retained only where a large activity difference suffices. 
SMILES strings, $K_i$ values, and pre-enumerated MMP lists were obtained from the accompanying repository.
\footnote{\url{https://github.com/MarkusFerdinandDablander/QSAR-activity-cliff-experiments}}

\subsection{Representations}

Six (embedding, metric) pairs configurations were computed using RDKit 2025.9.3: Morgan (radius 2, 1\,024 bits),\cite{rogers2010extended}
RDKit topological, and MACCS keys (166 bits),\cite{wildman1999prediction} each with Tanimoto and Dice similarity ($d = 1 - \text{sim}$). Three
learned embeddings were included: MolFormer\cite{ross2022large} -- version from the HuggingFace repository \texttt{ibm/MoLFormer-XL-both-10pct},
 ChemBERTa\cite{chithrananda2020chemberta} from the HuggingFace repository \texttt{DeepChem/ChemBERTa-77M-MLM}, and Chemeleon\cite{green2026deep}
(MPNN trained on Mordred descriptors;\cite{moriwaki2018mordred} obtained from the Zenodo repository \texttt{zenodo.org/records/15460715}). 
For learned embeddings, cosine distance was computed after mean-centering; $L^1$ and $L^2$ distances were scaled to $[0,1]$ at the 
90th percentile. All embeddings were held fixed; no fine-tuning was performed. 

\subsection{Activity Cliffs and Activity Gradients}

A pair $(i,j)$ constitutes an activity cliff when
\begin{equation}
    |\Delta\text{pK}_i| > A \;\text{ and }\; d(x_i, x_j) < s,
    \label{eq:cliff_binary}
\end{equation}
with $A = 1$ log unit throughout. This is equivalent to thresholding the pairwise activity gradient
\begin{equation}
    \mathcal{L}(i,j) = \frac{|\Delta\text{pK}_i|}{d(x_i,x_j)+\varepsilon}
\label{eq:sali}
\end{equation}
at $\mathcal{L} > A/s$, where $\varepsilon$ is the minimum resolvable non-zero distance per representation. The quantity 
$\mathcal{L}(i,j)$ is the Structure-Activity Landscape Index (SALI)\cite{guha2008structure} reinterpreted as the empirical Lipschitz constant 
of activity functional $\varphi$ under $d$: it measures the rate of activity change per unit of structural distance. Both the binary indicator and 
$\mathcal{L}(i,j)$ are explicit functions of $d$ and therefore of the representation.

\subsection{Six-Step Analysis Pipeline}

The six steps are ordered by geometric scale and logical dependence; each probes a property that the others cannot recover. Failure at an
earlier step renders later steps uninterpretable. Full diagnostic definitions and all numerical outputs are in the Supplementary Information (Tables~S1--S21).

\paragraph{Step 1: Pairwise distance geometry.}
We report: $p_0$ (fraction of zero-distance pairs), the BIC-selected
parametric family of the distance distribution, the coefficient of variation (CV; discriminative range), the relative contrast
RC $= (\bar{d}_{95} - \bar{d}_5)/\bar{d}_5$ (dynamic range relative to minimum resolution), and the hubness skewness $S_{N_k}$ at $k=5$
(neighbourhood reliability; $S_{N_k} \gtrsim 1.5$ flags hub-induced artefacts).

\paragraph{Step 2: Activity cliff enrichment.}
The cumulative cliff fraction $F(n)$ gives the fraction of all cliff pairs among the top $n\%$ most similar pairs.
For good representation the cumulative cliff fraction $F(n)$ should be as low as possible, 
approaching the diagonal from below, as it means that the fraction of cliffs among similar pairs is low. 
We fit $F(n) = n^s$ and derive the enrichment coefficient
\begin{equation}
    G = \frac{s-1}{s+1},
\label{eq:gini}
\end{equation}
with $G > 0$ indicating cliff depletion at high similarity (the favourable regime) and $G \approx 0$ indicating 
that nearest neighbours are indifferent to activity, i.e. the representation is not able to distinguish 
between similar and dissimilar pairs. The inflection steepness $k$ of the instantaneous cliff rate $f(n)$ is 
reported as a secondary diagnostic.

\paragraph{Step 3: Activity gradient distribution.}
We compute $\mathcal{L}(i,j)$ for all pairs with $d > 0$ and fit the Kohlrausch--Williams--Watts (KWW) survival function
\begin{equation}
    S(r) = a\exp\!\left(-kr^b\right)
\label{eq:kww}
\end{equation}
to its empirical survival function. The shape parameter $b$ is the primary statistic: $b = 2$ (Rayleigh ceiling) indicates a light-tailed,
smooth landscape; $b < 2$ indicates excess large gradients beyond what a smooth landscape produces. The 95th-percentile gradient
$r_{95} = \bigl(-\ln(0.05/a)/k\bigr)^{1/b}$ provides a complementary tail severity measure.

\paragraph{Step 4: Persistent homology of the cliff subspace.}
A Vietoris--Rips filtration on cliff-involved molecules tracks connected components ($H_0$): 
long-lived components indicate geometrically 
distinct, well-separated cliff-prone molecule clusters. We report the mean persistence 
$\mu_\text{pers}$, and maximum persistence $p_\text{max}$, supported by $L^2$ persistence landscape 
norm $\|\bm{\lambda}^{H_0}\|$ (all preferable larger). 

\paragraph{Step 5: Geometric probes of representational structure.}
\label{sec:step5_methods}
Three classifiers are trained on distance $|e_i - e_j|$ between molecules $i$ and $j$ in embedding as geometric probes: 
logistic regression (LR) tests linear separability; XGBoost tests nonlinear local discriminability; 
the Siamese network tests latent geometric richness beyond explicit feature interactions. 
Two gap statistics characterize the embedding geometry:
\begin{equation}
\Delta_\text{lin} = \overline{\text{AUC}_\text{XGB} - \text{AUC}_\text{LR}},
\label{eq:delta_lin}
\end{equation}
\begin{equation}
\Delta_\text{arch} = \overline{\text{AUC}_\text{Siam} - \text{AUC}_\text{XGB}},
\label{eq:gaps}
\end{equation}
As the input for the model was the absolute embedding difference, we took the mean over metrics.

Molecules were split into train/validation/test (70/10/20\%) by scaffold\cite{bemis1996properties} strategy.
For Logistic Regression and XGBoost the validation and train sets were joined. Pairs were class-balanced by downsampling; cross-split pairs were excluded.

\paragraph{Step 6: Chemical ground through benchmarking.}

Two structurally defined sub-populations probe representational completeness independently of the pipeline's own similarity measure.
\textit{Stereoisomers} and \textit{Matched molecular pairs}.
For learned embeddings on both sub-populations, and for all representations on MMPs, 
we fit a Beta distribution to the non-zero scaled distances among cliff pairs and report 
the coefficient of variation $\text{CV} = \sigma/\mu$ as the primary ranking statistic.

\subsection{Implementation}
All the pipeline was implemented in Python 3.11 using RDKit 2025.9.3, \texttt{scipy}, \texttt{xgboost}, \texttt{PyTorch} (Siamese network), 
HuggingFace \texttt{transformers} (ChemBERTa, MolFormer), and Chemeleon inference script based on \texttt{chemprop} 2.2.0.

\section{Results}

The six pipeline steps are applied to fifteen (embedding, metric)
configurations across three activity datasets with distinct targets,
known for their activity cliffs challenges. 
We present them in order of pipeline steps,
emphasizing  what type of geometric information is contributed at each step.

\subsection{Step 1: Pairwise Distance Geometry}

\begin{figure}[htbp]
  \centering
  \includegraphics[width=\linewidth]{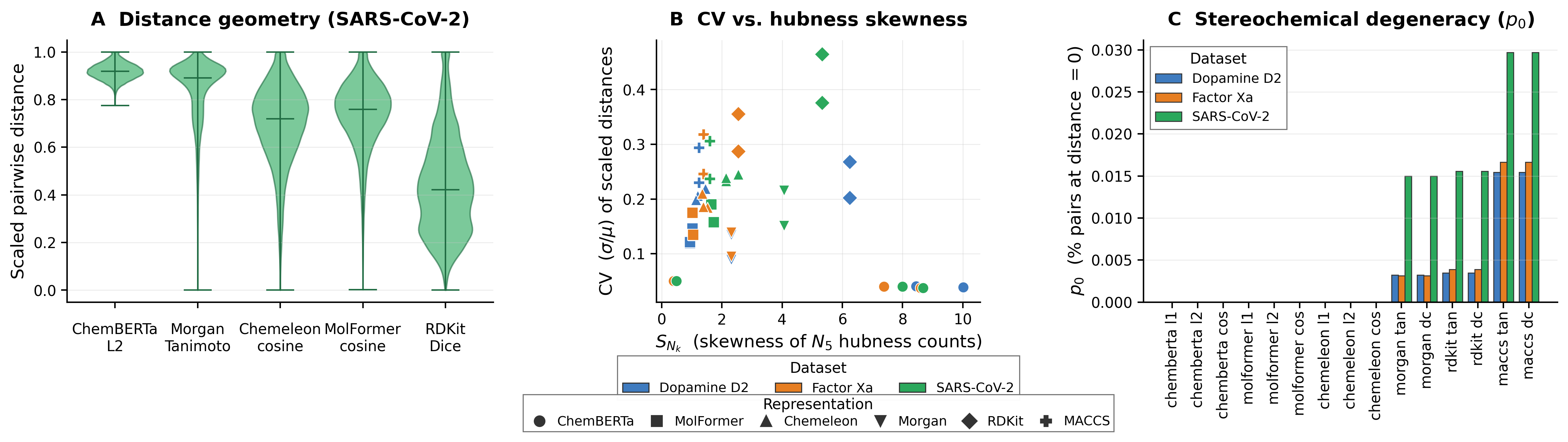}
  \caption{Pairwise distance geometry across all fifteen
    (embedding, metric) configurations. \textbf{A}: Scaled
    pairwise distance distributions on SARS-CoV-2 for five
    representative configurations, showing three qualitatively
    distinct regimes: embedding collapse (ChemBERTa), bimodal
    scaffold-structured geometry (Morgan Tanimoto), and
    continuous unimodal spread (Chemeleon cosine, MolFormer
    cosine, RDKit Dice). \textbf{B}: Coefficient of variation
    (CV) versus hubness skewness ($S_{N_k}$) for all fifteen
    configurations across all three targets. Configurations in
    the upper-left (high CV, low $S_{N_k}$) have both
    sufficient discriminative range and reliable neighbourhood
    structure; ChemBERTa occupies the low CV values
    regardless of metric. \textbf{C}: Fraction of zero-distance
    pairs $p_0$ per configuration, demonstrating that
    stereochemical degeneracy is a property of the fingerprint
    family, not of the metric. Full numerical values are
    reported in Tables~S1--S3.}
  \label{fig:step1_geometry}
\end{figure}

As a first step in characterizing the geometry induced by each (embedding, metric) pair, we
analyzed the distribution of pairwise molecular distances across the dataset. This analysis
serves a dual purpose: it identifies fundamental representational limitations that constrain
all subsequent cliff-based analyses, and it provides a set of geometry-level diagnostics that
can be interpreted independently of bioactivity data. In particular, we measure the fraction 
of zero-distance pairs $p_0$ - the pairs of structurally unresolved molecules. We also characterize
the discriminative range by coefficient of variation (CV, measuring global discriminative range) 
and median relative contrast (RC, measuring local discriminative range). We also 
fit the probability distribution selected by Bayesian Information Criterion (BIC) to the distance distribution. 
Finally, to investigate, if the dataset is balanced, so there are no hubs - molecules similar to many other
molecules, potentially artificially inflating cliff fraction, we measure the hubness k-nearest 
neighbors of molecules, expressed as the skewness of the $k$-occurrence distribution at $k=5$.
The full description of the diagnostics is in the Supplementary Information along with the numerical 
values (Tables~S1--S3). Figure~\ref{fig:step1_geometry} shows the key patterns.

\textbf{ChemBERTa} produces severely concentrated distances across all
metrics and targets (CV $\approx 0.04$--$0.05$; $S_{N_k}$ up to
$10.0$ under $L^1$/$L^2$ - compare Figure~\ref{fig:step1_geometry}A and B). The apparent low hubness under cosine
($S_{N_k} \approx 0.43$--$0.49$) is an artefact of concentration, not
genuine neighbourhood structure. This means that the embedding space is geometrically
degenerate with most of the molecules being similar to each other.

\textbf{Fingerprints} offer wide discriminative range but carry a hard
cost -- number of unresolvable pairs $p_0 = 0.003$--$0.030$ across all types 
(compare Figure~\ref{fig:step1_geometry}C). In particular, all
stereoisomeric pairs are assigned distance zero. RDKit achieves the
highest coefficient of variation CV (up to $0.47$) but also the highest hubness ($S_{N_k}$ up
to $6.2$ on Dopamine D2 - Figure~\ref{fig:step1_geometry}B), which may artificially inflate the number
of cliff pairs. MACCS combines competitive range with the
most reliable neighbourhoods ($S_{N_k} \leq 1.60$). Morgan's distances
follow a Beta distribution uniquely among all configurations,
reflecting bimodal scaffold-based organization.

\textbf{Chemeleon and MolFormer} produce $p_0 = 0$ and geometrically
sound spaces, but differ in stability: Chemeleon achieves higher CV
and relative contrast RC with cosine distance, while MolFormer is more isotropic ---
lower but consistent across all three targets and metrics
($S_{N_k} = 0.94$--$1.06$ on Factor Xa and Dopamine D2 - Figure~\ref{fig:step1_geometry}B). 

Step~1 does not rank representations --- it determines which ones are geometrically viable 
and what kind of space each defines. ChemBERTa fails this threshold entirely; all fingerprint 
and learned embedding configurations pass, but encode chemical diversity at qualitatively 
different scales: Morgan organises space around scaffold identity (bimodal Beta geometry), 
MACCS and RDKit at the fragment level (unimodal, wide range), and Chemeleon and MolFormer 
continuously (no zero-distance pairs, metric-dependent spread).

\subsection{Step 2: Activity Cliff Enrichment}
\begin{figure}[htbp]
  \centering
  \includegraphics[width=\linewidth]{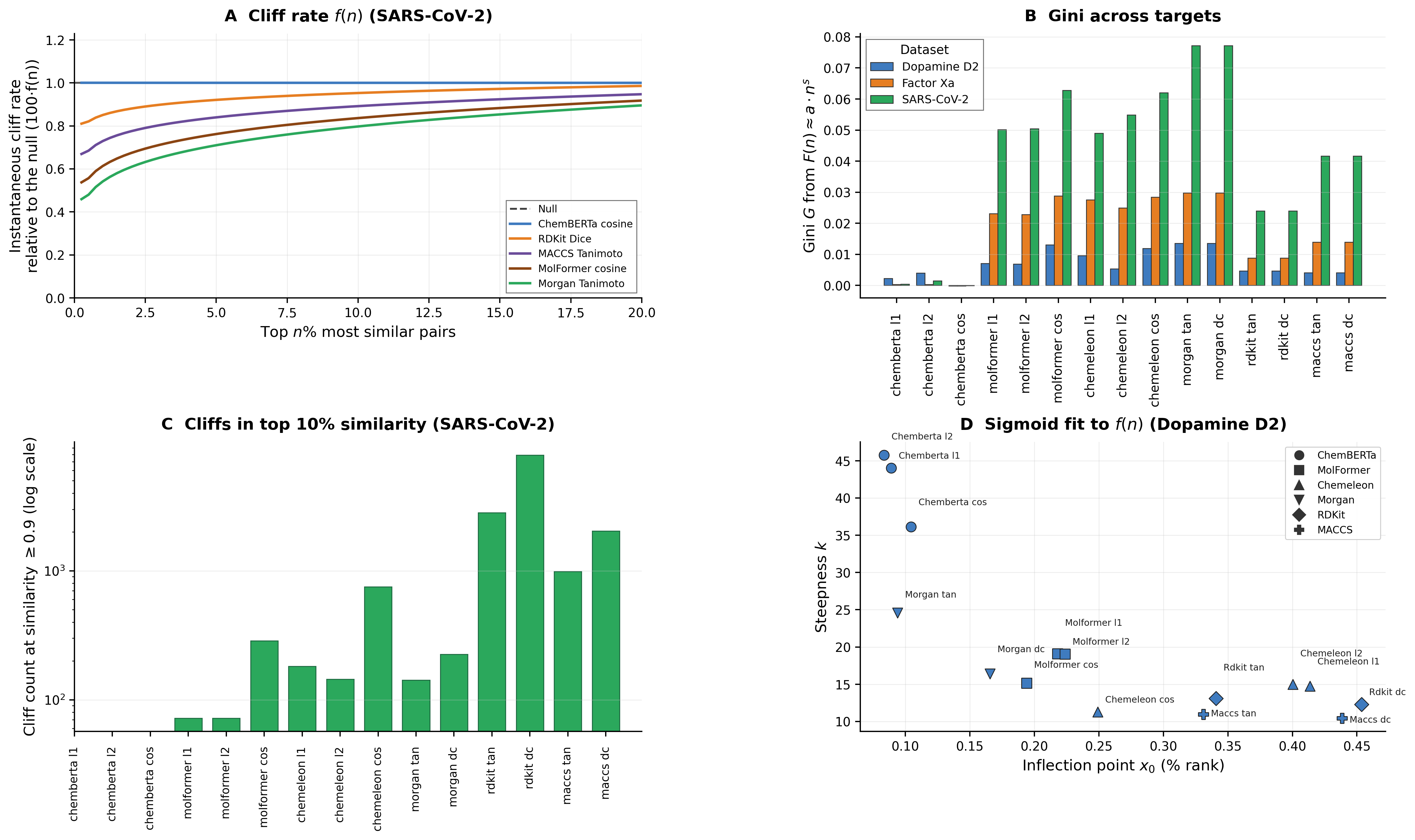}
  \caption{Activity cliff enrichment across all fifteen
    (embedding, metric) configurations.
    \textbf{A}: The instantaneous cliff rate $f(n)=\frac{dF(n)}{dn}$ as a function of the top $n\%$ most similar pairs. 
    Plotted is the curve multiplied by 100 for better readability. The null hypothesis is the value at $1$.
    Any value below $1$ denotes depletion of activity cliffs from the most similar pairs 
    to be on the same scale as the Lorenz enrichment curves --- the favourable regime in which structurally 
    similar compounds tend to have similar activity. ChemBERTa lies on exactly on the null hypothesis
    confirming that its metric space is indifferent to activity
    differences among nearest neighbours.
    \textbf{B}: $G$ values for all configurations across all
    three targets. The consistent ordering and monotonic
    compression across targets reveals that the decrease in
    maximum achievable $G$ from SARS-CoV-2 to Dopamine D2 is
    a property of the target, not of any individual representation.
    \textbf{C}: Number of cliff pairs identified at 90\%
    structural similarity on SARS-CoV-2 (log scale). The
    55-fold range across configurations (142 to 7\,903) is
    driven entirely by representation geometry, not by biology. 
    \textbf{D}: The sigmoid parameters $k$ vs $x_0$ for all configurations on Dopamine D2.
    Representations with high $G$, low $k$, and high $x_0$ are the best 
    for cliff prediction. Yet for all representations in bottom-right corner the corresponding 
    $G$ values are rather low. Full enrichment statistics are in Tables~S4--S6.}
  \label{fig:step2_enrichment}
\end{figure}

Good distance geometry (Step~1) does not guarantee that cliff pairs
will in fact be well separated. The crucial diagnostic is the enrichment 
(or rather depletion) curve $F(n)$ -- the number of activity cliffs among 
the top $n\%$ most similar pairs -- or its derivative $f(n)$, the instantaneous cliff rate 
(Figure~\ref{fig:step2_enrichment}A). If the structural similarity is a good predictor of the
activity similarity, the number of cliffs should be lower than random, i.e. $F(n) < n/100$. In the
instantaneous cliff rate this means, that $f(n) < 1/100$ (Figure~\ref{fig:step2_enrichment}A) and the 
lower the curve is the better (less cliffs for similar molecules). This intuition might be quantified 
by the enrichment coefficient $G$ (Eq.~\ref{eq:gini}, Figure~\ref{fig:step2_enrichment}B - the higher 
the better).

In particular, \textbf{ChemBERTa} produces $G \approx 0$ on all targets ---
cliff pairs are distributed as randomly as non-cliff pairs. On the other hand,
\textbf{Morgan Tanimoto} achieves the highest and most consistent $G$
across all three targets ($0.077$, $0.030$, $0.014$, Fig.~\ref{fig:step2_enrichment}B), mechanistically
consistent with its bimodal geometry: scaffold clustering places
structural neighbours at exactly the short-distance end where cliff
partners should appear. The impact is stark: at 90\% similarity,
Morgan Tanimoto identifies (only) 142 cliff pairs on SARS-CoV-2, while Chemeleon cosine
identify 752, while RDKit Dice 7\,903 respectively --- a 55-fold range driven
by representation geometry, not biology (Figure~\ref{fig:step2_enrichment}C). 
Among learned embeddings, MolFormer cosine and Chemeleon cosine lead ($G \approx 0.063$ and
$0.062$ on SARS-CoV-2); cosine consistently outperforms $L^1$/$L^2$
within both families. RDKit, despite the highest discriminative range
in Step~1, produces the lowest $G$ among fingerprints --- wide
distance spread, potentially favoring cliff identification, does not 
automatically imply cliff enrichment.

The Gini coefficient $G$ captures the overall magnitude of cliff
depletion but not its distribution across the similarity spectrum.
The parameters $k$ (steepness) and $x_0$ (inflection point) of the sigmoid fit to the 
instantaneous cliff rate $f(n)$ function (see Supplementary Information 
Section S2.4 for details) reveal a complementary dimension: 
the \emph{range} over which depletion is operative.
On the Dopamine D2 dataset (Tables~S6, Fig.~\ref{fig:step2_enrichment}D), 
this distinction is particularly informative.
Morgan Tanimoto achieves the highest $G$ ($0.0135$) yet also the highest
$k$ ($24.6$) and lowest $x_0$ ($0.094$) among non-degenerate
configurations. This combination means that cliff depletion, while
globally strong, is concentrated within an extremely narrow band of the
most similar pairs: beyond the top $\sim 0.1\%$ most similar compound
pairs, cliff frequency rises sharply toward the null baseline. The
bimodal scaffold-based geometry identified in Step~1 provides the
mechanistic explanation --- within-scaffold pairs cluster at very short
distances (producing strong depletion), but the transition to the
cross-scaffold population is abrupt, so the cliff-free regime collapses
quickly as soon as any cross-scaffold pairs are included.

In contrast, MACCS Dice achieves a lower $G$ ($0.0041$) but a much
higher $x_0$ ($0.439$) and lower $k$ ($10.4$): weaker depletion,
but distributed across a substantially wider similarity range. RDKit
Dice shows the same pattern ($x_0 = 0.454$). These representations
maintain partial activity consistency across a broad neighbourhood ---
not just at the most stringent similarity threshold. MolFormer cosine
and Chemeleon cosine occupy an intermediate position ($x_0 \approx
0.19$--$0.25$, $k \approx 11$--$15$): moderate depletion that is
neither as concentrated as Morgan nor as broadly distributed as
fingerprints with Dice similarity.

The practical consequence for representation selection depends on the
similarity threshold used in the downstream application. At very
high similarity thresholds ($\geq 90\%$), Morgan Tanimoto's narrow
but strong depletion makes it the best choice. At moderate thresholds
($70$--$85\%$), the broader depletion of MACCS or RDKit with Dice
similarity becomes comparatively more useful. A representation with
both high $G$ and high $x_0$ --- strong depletion across a wide
similarity range --- would be ideal but is not observed in the present
benchmark, identifying a geometric property that current molecular
representations do not simultaneously achieve.

Interestingly, the maximum achievable $G$ decreases monotonically from SARS-CoV-2
to Factor Xa to Dopamine D2 for every representation simultaneously,
signalling a target-level landscape property, studied quantitatively in Step~3.

\subsection{Step 3: Activity Gradient Distribution}

\begin{figure}[htbp]
  \centering
  \includegraphics[width=\linewidth]{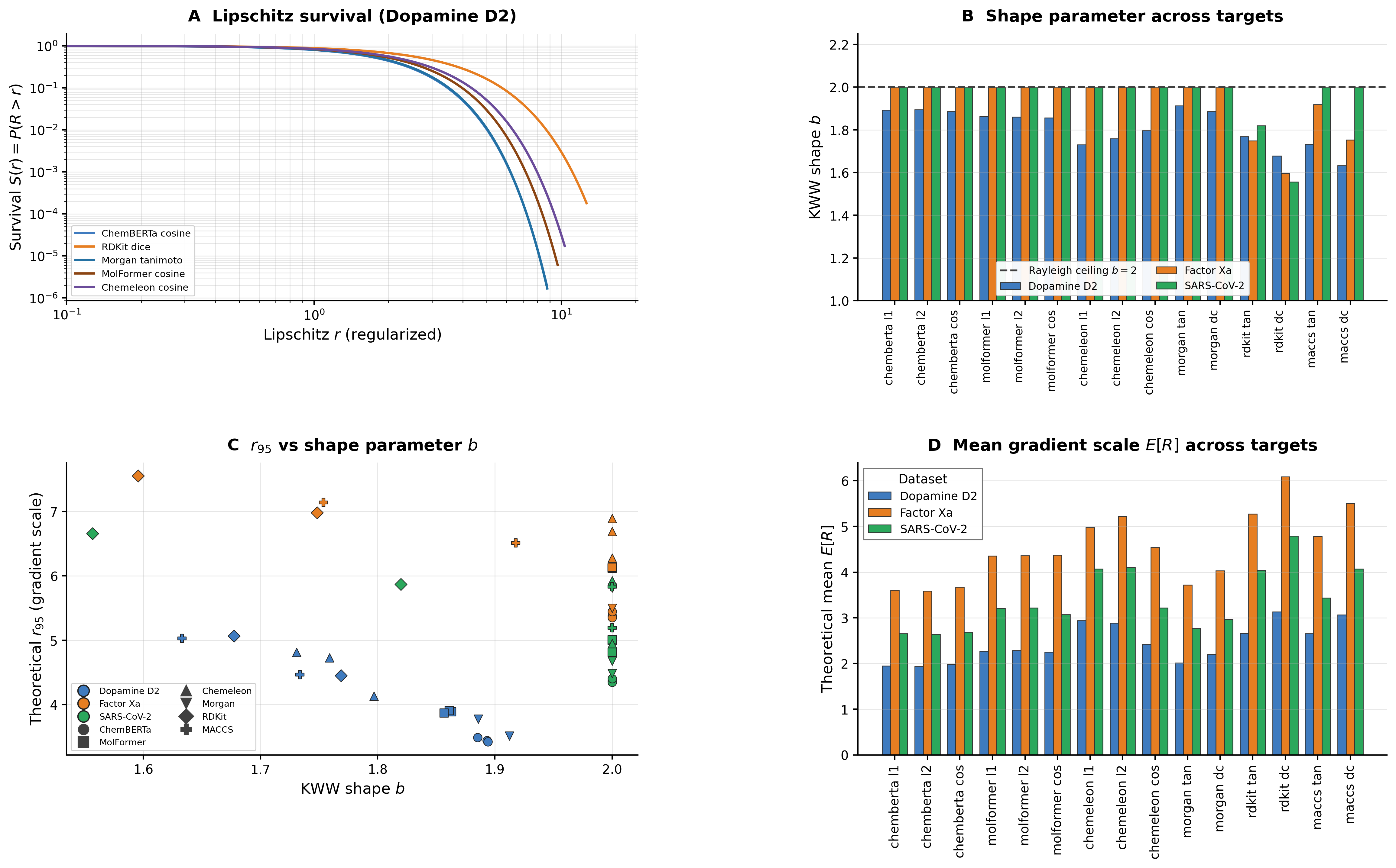}
  \caption{Activity gradient distribution diagnostics (Step 3)
    across all fifteen (embedding, metric) configurations.
    \textbf{A}: KWW survival function fits on Dopamine D2 for
    representative configurations. Configurations with faster decay are characterized by 
    larger value of shape parameter $b$, seemingly denoting smoother landscape with less 
    frequent extreme activity gradients. Longer decay with $b<2$ indicates a heavier tail
    (stretched exponential) with disproportionately frequent
    extreme activity gradients. \textbf{B}: Shape parameter $b$ of the KWW fit to the survival function
    across all configurations and targets. The dashed line marks
    the Rayleigh ceiling $b=2.00$ (Gaussian random landscape). No configuration reaches this
    ceiling on Dopamine D2, while most do on SARS-CoV-2 ---
    a target-level roughness signature independent of
    representation. \textbf{C}: Joint $r_{95}$ vs $b$ scatter.
    Points in the top-left region (low $b$, high $r_{95}$)
    indicate the roughest configurations; the systematic
    separation of Dopamine D2 points confirms the target-level
    roughness finding from Panel B. \textbf{D}: Mean gradient $\mathbb{E}[R]$ measured for all configurations. 
    The smaller values indicate smoother landscape on average - smaller jumps in activity for two randomly chosen 
    similar molecules. Full KWW parameters are in Tables~S7--S9.}
  \label{fig:step3_gradients}
\end{figure}

Replacing the binary cliff indicator with the continuous gradient
$\mathcal{L}(i,j)$ (Eq.~\ref{eq:sali}) reveals properties invisible
to Steps~1--2. These might be characterized by the measuring the distribution of the activity
gradient $\mathcal{L}(i,j)=\frac{|\Delta\text{pK}_i|}{d(x_i,x_j)+\varepsilon}$.
In particular, one may ask for the (statistical) probability this value exceeds a given threshold $r$.
This is the survival curve $S(r)$ (Fig.~\ref{fig:step3_gradients}A), which may be 
well fitted by the KWW distribution (Eq.~\ref{eq:kww}) with 
the shape parameter $b$ and the decay rate $k$ (see Supplementary Information Section S3.1--S3.4
for details). The shape parameter $b$ characterizes the tail shape of the distribution, i.e. how 
probable are the extreme activity gradients. In particular, $b=2$ (fast decay) corresponds to the Rayleigh 
ceiling observed when the extreme activity gradients are as frequent as a Gaussian random field 
would produce (see Supplementary Information Section S3.1--S3.4). The lower values indicate more than
random extreme activity gradients.

On \textbf{SARS-CoV-2}, eleven of fifteen configurations reach the
Rayleigh ceiling $b = 2.00$ (light-tailed, smooth landscape) -- Fig.~\ref{fig:step3_gradients}B. The
exceptions are RDKit Tanimoto ($b = 1.82$) and RDKit Dice ($b = 1.56$)
--- the widest distance spread in Step~1 and moderate enrichment in
Step~2, yet the heaviest gradient tails here. 

On \textbf{Factor Xa}, learned embeddings and Morgan retain $b = 2.00$
while MACCS and RDKit fall below (RDKit Dice: $b = 1.60$,
$r_{95} = 7.55$ vs $5.81$ for Morgan Tanimoto) --- a roughness
invisible to the binary enrichment analysis of Step~2.

On \textbf{Dopamine D2}, no configuration reaches $b = 2.00$
(maximum $b = 1.91$, Morgan Tanimoto). 
The universal $b<2$ on Dopamine D2 indicates that no representation reduces the activity gradient 
distribution to the Gaussian baseline — structured discontinuities persist in the landscape 
regardless of how chemical space is embedded, identifying that Dopamine D2 activity landscape is an intrinsically 
rougher (more prone to extreme activity gradients, and therefore cliffs) than SARS-CoV-2 and Factor Xa.
This may be due to selective binding pocket.

The shape parameter $b$ tells how much of the SALI-extreme values are in the tail of the distribution,
but it does not tell how large these values are. This is quantified by the 95th-percentile gradient 
$r_{95}$ indicating how large the extreme SALI values are.
The configurations with small $b$ and large $r_{95}$ are the roughest and most challenging landscape for 
cliff prediction. When plotting jointly $b$ and $r_{95}$ (Fig.~\ref{fig:step3_gradients}C) one can see a clear
separation of the Dopamine D2 dataset. But, although much of the SALI-extreme values are in the tail of the
distribution for this dataset, the SALI values are rather small compared to other datasets 
(low $r_{95}$ values). On the other hand,
on Factor Xa RDKit Dice has low $b=1.5955$, but high $r_{95}=7.5502$ (Tab. S8) meaning, that in this particular 
representation there are much more pairs with large SALI values, again, independently of the biology of the target.

Interestingly, the order of the target landscape roughness following from the analysis of the shape parameter $b$ 
is not reflected in the order of the mean gradient $\mathbb{E}[R]$ (Fig. \ref{fig:step3_gradients}D). In particular,
the highest values of $\mathbb{E}[R]$ are observed for Factor Xa protein, and lowest for Dopamine D2 receptor. This 
indicates, that taking randomly two similar molecules from Factor Xa dataset, the activity difference between them is 
on average highest. However, both $b$ and $\mathbb{E}[R]$ are strongly representation-dependent, 
reflecting that landscape roughness —- whether measured by tail severity or mean gradient 
magnitude —- is a property of the (embedding, metric) pair as much as of the target.
 
Step~4 examines whether this roughness has a corresponding topological structure in the cliff subspace.

\subsection{Step 4: Persistent Homology of the Cliff Subspace}

\begin{figure}[htbp]
  \centering
  \includegraphics[width=\linewidth]{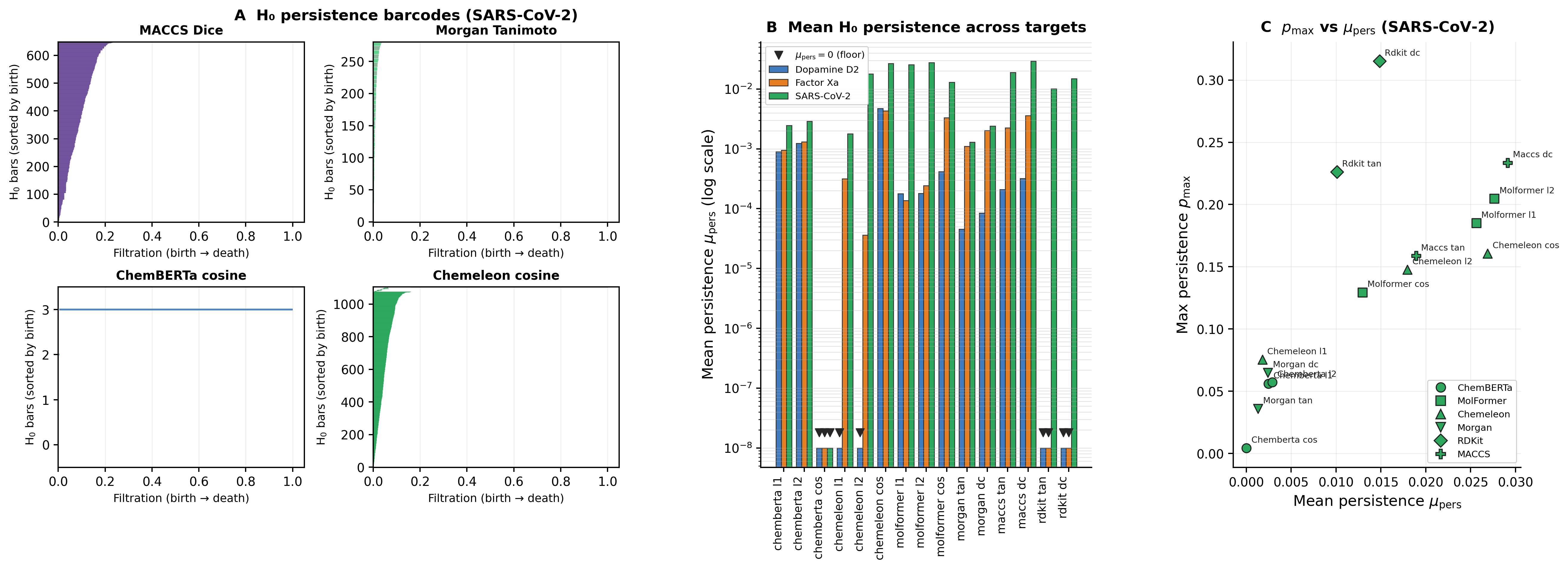}
  \caption{Persistent homology of the activity cliff subspace
    (Step~4) across all fifteen (embedding, metric) configurations.
    \textbf{A}: $H_0$ persistence diagrams for four
    representative configurations on SARS-CoV-2. Long-lived
    components (long bars) indicate that
    cliff clusters remain topologically separated across a wide
    range of filtration scales; the complete collapse of
    ChemBERTa cosine (only one long bar) is confirmed
    by a topological route independent of Steps~1--3.
    \textbf{B}: Mean persistence $\mu_\text{pers}$ (log scale)
    reveals topological collapse of the cliff subspace (denoted by triangle markers).
    \textbf{C}: Maximum persistence $p_\text{max}$ vs mean persistence $\mu_\text{pers}$
    on SARS-CoV-2. The points in the top-right corner (large $\mu_\text{pers}$, large $p_\text{max}$)
    are the richest and most interpretable landscape for medicinal chemistry, as there are
    many cliff clusters of comparable isolation. 
    Full data in Tables~S10--S12.}
  \label{fig:step4_homology}
\end{figure}

Pairwise statistics analyzed in previous steps cannot capture how cliff-involved molecules are
collectively organised across the full embedding space. Even if the cliff-prone molecules are pairwise far apart,
they may still be globally clustered together. As a result, their mutual disctance may not be 
a good predictor of the cliff presence. This phenomenon is naturally studied by tracking the 
connected components ($H_0$) in persistent homology. Due to its construction (see section S4.1 in the 
Supplementary Information for details), the components group together cliff-prone molecules with mutual distance 
less than the filtration parameter $\varepsilon$. The longer the components are present, as the function of the
filtration parameter $\varepsilon$, the more separated are the molecule clusters.

Usually, the lifespan of the components is presented in a form of diagrams (Fig.~\ref{fig:step4_homology}A). 
Alternatively, the collective lifespan of the components can be characterized by mean persistence $\mu_\text{pers}$, 
(Fig.~\ref{fig:step4_homology}B) or maximum persistence $p_\text{max}$ of the bars.

The $\mu_\text{pers}$ approaching zero (topologicalcollapse) means all cliff-involved molecules are in the 
same region of the embedding space. This can be either due to fundamental metric degeneracy (evidenced in Step~1) 
or target-specific compression.  In particular, ChemBERTa cosine collapses on all three targets 
(Fig.~\ref{fig:step4_homology}B) -- the effect of global distance concentration (Step~1). More interestingly, 
RDKit Tanimoto and Dice collapse on Factor Xa and Dopamine D2, while Chemeleon $L^1$/$L^2$ collapse on 
Dopamine D2 only (Fig.~\ref{fig:step4_homology}B). This means, that the global scaffold of cliff-prone molecules may 
be easy to identify in these representations, which might be of use for medicinal chemistry. 

In general, the $\mu_\text{pers}$ can be used to characterize the representations, in which the cliff-prone 
molecules form distinct, well-separated clusters. These are better for structural biologists 
working on mechanism understanding, as each cluster corresponds to a potentially different pharmacological 
explanation —- a different binding mode transition, a different stereocentre sensitivity, 
a different transformation-sensitive pharmacophore. 

The mean persistence $\mu_\text{pers}$ and maximum persistence $p_\text{max}$ are strongly correlated 
(Fig.~\ref{fig:step4_homology}C). However, in some exceptional cases one can observe small $\mu_\text{pers}$ with 
large $p_\text{max}$. This can be seen for RDKit Tanimoto/Dice on SARS-CoV-2. One or a small number of coherent 
cliff-prone scaffold families well-separated from an otherwise undifferentiated background. This is the most 
actionable finding for a medicinal chemist: a specific scaffold is identifiable as cliff-prone and can be 
flagged or avoided, while the rest of the cliff population remains unpredictable.

One can also see, that the mean persistence $\mu_\text{pers}$ for a given representation is usually the largest for
SARS-CoV-2 and comparable for Factor Xa and Dopamine D2. This possibly stems from the fact, that SARS-CoV-2 PostEra 
Moonshot dataset is extraordinarily chemically diverse by design, while Factor Xa and Dopamine D2 tend to 
concentrate around a smaller number of established pharmacophores.

In general, the mean persistence $\mu_\text{pers}$ can be used to quantify the ability to separate the cliff-prone
molecules by distance. In particular, the cliff prediction based on the distance between two molecules should be 
easier for representations with large $\mu_\text{pers}$ and almost impossible for representations with
topological collapse. Showing this quantitatively is the aim of Step~5.

\subsection{Step 5: Geometric Probes of Representational Structure}

We trained three classifiers (logistic regression, XGBoost and Siamese network) 
on the absolute embedding difference $|e_i - e_j|$ between the molecules $i$ and $j$ and tested their 
performance on the scaffold split. This tests, if the geometric
properties of Steps~1--4 translate to supervised signal. Full AUC
values and gap statistics are in Tables~S13--S15. As in each case the input for the model was
the absolute embedding difference, it was dependent only on the representation and not on the metric. 
In what follows we present the mean ROC AUC values over each metric.

\begin{figure}[htbp]
  \centering
  \includegraphics[width=\linewidth]{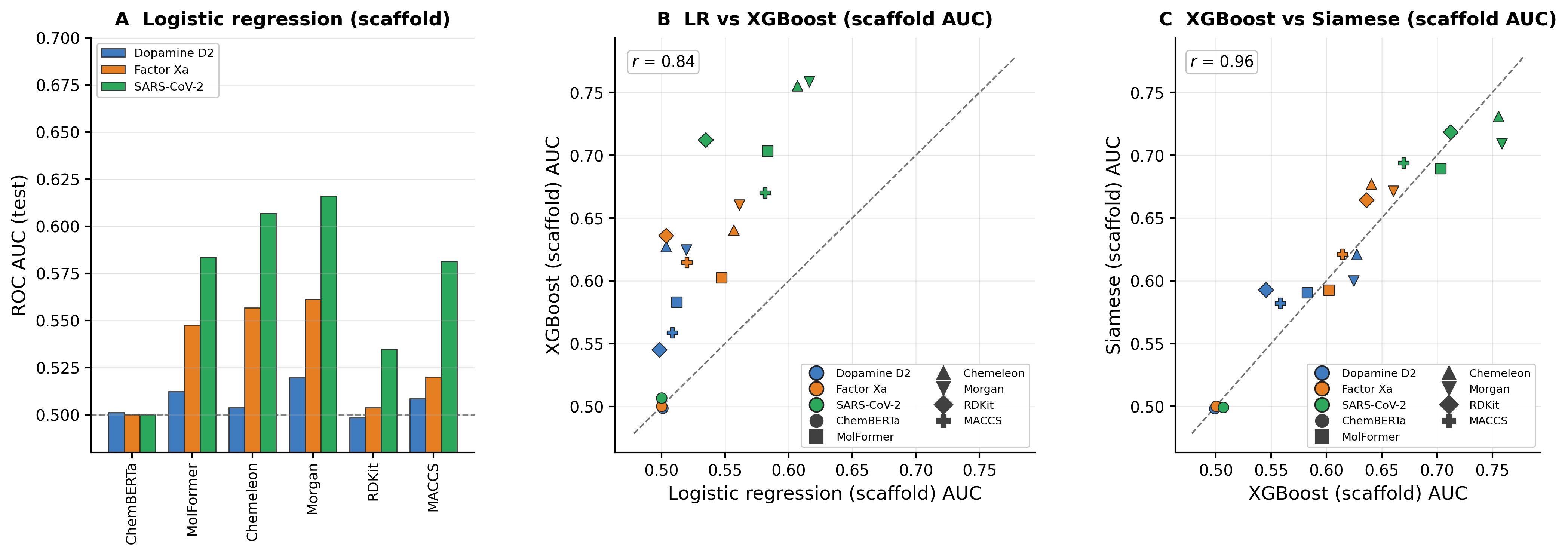}
  \caption{Predictive benchmarking results (Step~5) across all
    six embeddings averaged over metrics.
    \textbf{A}: ROC AUC values for the scaffold split in logistic regression for all 
    embeddings and all three targets. The ROC AUC values are close to random ($\approx 0.5$) 
    for the configurations falling to topological collapse (Step~4).
    \textbf{B}: XGBoost vs Logistic Regression performance on scaffold split measured by ROC AUC. 
    For all representations (excluding ChemBERTa) the XGBoost performance is better than the 
    Logistic Regression showing, that the identification of the cliffs is a non-linear problem. 
    The difference between the XGBoost and Logistic Regression performance is 
    denoted as $\Delta_\text{lin}$ (Eq.~\ref{eq:delta_lin}).
    \textbf{C}: Architecture gap $\Delta_\text{arch}$ heatmap.
    The sign of $\Delta_\text{arch}$ separates fingerprints
    (negative; informationally saturated for tree methods) from
    continuous embeddings (positive; latent structure exploitable
    by learned re-encoding). Full AUC data and gap statistics in
    Tables~S13--S15.}
  \label{fig:step5_prediction}
\end{figure}

Consistently with the results of previous steps, the ROC AUC values indicate, that 
different datasets have different difficulty level. In particular, as the Dopamine D2 dataset,
characterized by the most roughest landscape (Step~3) is the most difficult for all classifiers 
(Fig.~\ref{fig:step5_prediction}A). On the other hand, as predicted, the configurations falling to 
topological collapse (Step~4) cannot extract any meaningfull signal from the distance alone, so their ROC AUC is 
close to random (Fig.~\ref{fig:step5_prediction}A) - the ROC AUC values are $\approx 0.5$ for 
ChemBERTa on all datasets, RDKit on Dopamine D2 and Factor Xa, while Chemeleon
on Dopamine D2 only slightly better than random.

Moreover, for ChemBerta the metric does not increase when changing the model to non-linear XGBoost
(Fig.~\ref{fig:step5_prediction}B). For other models the difference between the XGBoost and 
Logistic Regression performance ($\Delta_\text{lin}$) is positive showing that the 
cliff identificationproblem is nonlinearly structured in every evaluated embedding space. Yet, the
gain is roughly proportional to the Logistic Regression performance and does not change much the order
of the representations nor targets.

The more interesting results are obtained when analyzing the architecture gap $\Delta_\text{arch}$ 
(Fig.~\ref{fig:step5_prediction}C). It compares the result of a Siamese network, which has an access 
to both embeddings, with the result of XGBoost, which operates on the absolute embedding difference.
Therefore, in principle, the Siamese network can restore some signal which is lost in the absolute 
difference. And indeed, a small positive gain is observed for the fingerprint representations, 
where each bit denotes the presence or absence of a specific substructure.

The effect is not visible in case of Morgan Fingerprints, where $\Delta_\text{arch}$ are slightly
negative (symbols in Fig.~\ref{fig:step5_prediction}C below diagonal). This is especially evindent in 
the case of SARS-CoV-2 dataset. But in this case most of the representations do not gain from the
change of the model architecture. It shows, that the results obtained with a simple XGBoost model
are already better than the results which are obtained with an unoptimized Siamese network. 
On the other hand, for harder datasets like Dopamine D2 and Factor Xa the gain is more visible.

The exception is, again, ChemBERTa, where there is no gain at all, which proves, that not only the
distance is not informative enough, but the compound representations are lying in the same region 
of the embedding space, as already shown in Step~4.

\subsection{Step 6: Chemical ground truth benchmarking}
\label{sec:external}
The five pipeline steps characterize the activity landscape using the
internal similarity structure induced by each (embedding, metric) pair:
distance distributions, cliff enrichment curves, gradient survival
functions, and topological organization are all computed relative to
the distances that the representation itself assigns. 

The last step inverts this logic. Rather than asking how the
representation organizes the full compound library, it asks a sharper
and more local question: for molecular pairs that are known to be
structurally related by an explicit, representation-independent
criterion and that exhibit a large activity difference, what distance
does each representation assign? These pairs are similar by definition
--- not by the representation's own metric --- and their activity
difference is known to be driven by a specific, chemically
interpretable structural feature. A representation that assigns large
and discriminative distances to such pairs is one whose notion of
structural proximity is sensitive to precisely the changes that matter
for activity. One that collapses them to zero or assigns uniformly
small distances is one that is blind to the structural features
responsible for the cliff, regardless of how well it performs on the
internal pipeline diagnostics.

Two structurally defined sub-populations are examined. \textit{Stereoisomers}
--- compound pairs sharing an identical molecular graph but differing
in the spatial arrangement of atoms --- represent the most extreme
possible test of representational completeness: the relevant structural
difference is entirely three-dimensional, invisible to any
graph-based encoding. \textit{Matched molecular pairs (MMPs)} ---
compounds related by a single, well-characterized chemical
transformation applied to a shared core --- represent a richer and
more diverse probe of transformation-level sensitivity, where the
activity-driving structural change is explicit and chemically
interpretable. In both cases, the primary ranking statistic is the
coefficient of variation (CV) of the Beta-fitted distance distribution
among cliff pairs within the sub-population: larger CV indicates that
the representation assigns more variable and discriminative distances
to pairs whose activity difference is known to be structurally driven,
reflecting greater sensitivity to the specific features at stake.
Together, the two analyses expose complementary geometric limitations
that the internal pipeline cannot resolve --- stereochemical blindness
and transformation-level insensitivity --- and provide a direct
empirical test of the representation's faithfulness to the chemistry
responsible for activity cliffs.

\begin{figure}[htbp]
  \centering
  \includegraphics[width=\linewidth]{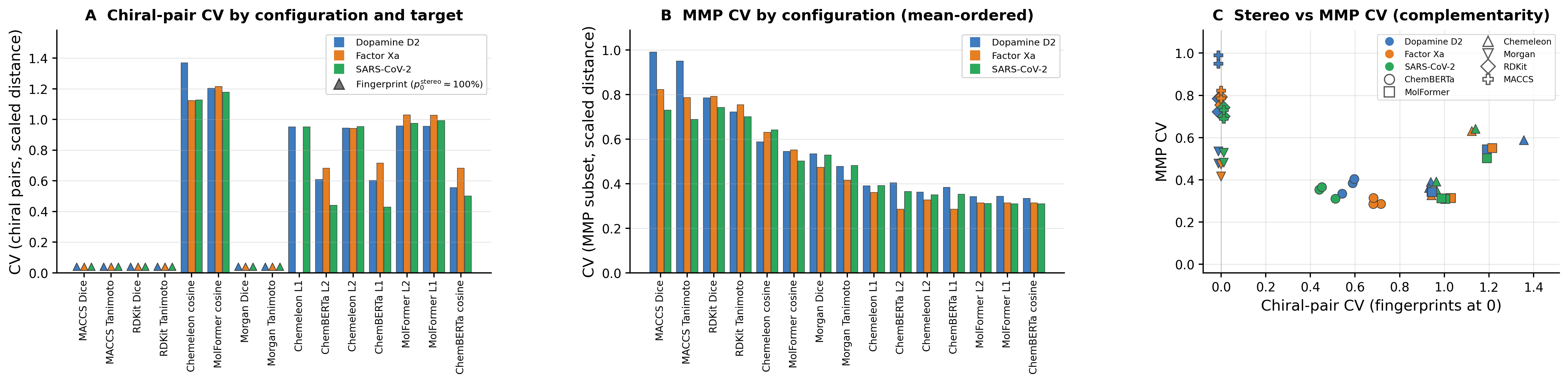}
  \caption{Stereoisomer cliff pair analysis across all fifteen
    (embedding, metric) configurations.
    \textbf{A}: Coefficient of variation (CV) across all learned
    embedding configurations and all three targets. MolFormer
    cosine achieves the most consistent CV across targets
    (range $0.037$); Chemeleon cosine achieves the highest
    single-target CV (Dopamine D2: $1.368$) but with greater
    cross-target variation (range $0.245$). Full Beta fit
    parameters in Tables~S16--S18.}
    \textbf{B}: MMP CV for all configurations and targets.
    Fingerprint representations with Dice similarity dominate
    the high-CV tier; transformer embeddings with $L^1$/$L^2$
    metrics occupy the low-CV tier. Unlike the stereoisomer
    analysis, no configuration is fully degenerate --- all
    produce valid Beta fits. Full Beta fit
    parameters in Tables~S19--S21.
    \textbf{C}: Stereo CV vs MMP CV scatter for all
    configurations. No representation simultaneously achieves
    high sensitivity on both axes. The plot directly visualises
    the complementarity between fingerprints (high MMP CV,
    zero stereo CV) and learned embeddings with cosine distance
    (high stereo CV, intermediate MMP CV). 
  \label{fig:step6_stereo_and_mmps}
\end{figure}

\subsubsection{Stereoisomers}

Stereoisomeric cliff pairs share an identical molecular graph and
differ only in spatial arrangement; their activity difference is
driven entirely by three-dimensional pharmacophore complementarity.
Full data are in Tables~S16--S18.

All six \textbf{fingerprint} configurations assign $p_{0,\text{stereo}}
= 100\%$ on every target without exception (Figure~\ref{fig:step6_stereo_and_mmps}A) --- 
a hard architectural failure. The SALI gradient $\mathcal{L}(i,j)$
is undefined for every stereoisomeric pair under any fingerprint metric.

Among learned embeddings, \textbf{MolFormer cosine} achieves the most
stable CV across all three targets (CV $= 1.178$, $1.215$, $1.202$;
cross-target range $0.037$). \textbf{Chemeleon cosine} reaches the
highest single-target CV (Dopamine D2: $1.368$) but with six times
wider cross-target variation (range $0.245$), indicating
target-dependent rather than stable stereochemical sensitivity.
\textbf{ChemBERTa} assigns non-zero distances ($p_{0,\text{stereo}}
= 0$) but with CV $= 0.428$--$0.716$ --- uniformly large distances
with little discrimination, the same concentration artefact seen in
Steps~1--5. Within both MolFormer and Chemeleon, cosine
consistently outperforms $L^1$ and $L^2$.

\subsubsection{Matched Molecular Pairs}

MMP cliff pairs are related by a single explicit chemical
transformation on a shared core. Full data are in Tables~S19--S21;

The MMP ranking is the near-exact \textbf{inverse of the stereoisomer
result} (Figure~\ref{fig:step6_stereo_and_mmps}B). MACCS Dice leads across all three targets (CV $= 0.730$,
$0.822$, $0.990$), followed closely by RDKit Dice ($0.742$, $0.792$,
$0.784$). The fingerprints that failed categorically for stereoisomers
dominate here; learned embeddings that led for stereoisomers fall
behind. The inversion reflects structural scale: MACCS keys encode
fragment-level patterns at exactly the scale at which single-step
MMP transformations manifest.

Unlike stereoisomers, \textbf{no configuration assigns distance zero}
to MMP pairs; the CV range is graded ($0.285$--$0.990$, 3.5-fold).
The low end is occupied by ChemBERTa and MolFormer $L^1$/$L^2$
(CV $\approx 0.29$--$0.34$), which assign nearly uniform distances
regardless of transformation type. Within each family, Dice outperforms
Tanimoto ($+0.03$--$+0.05$ CV) and cosine outperforms $L^1$/$L^2$
($+0.19$--$+0.28$ CV) across all three targets.

The stereo-vs-MMP scatter (Figure~\ref{fig:step6_stereo_and_mmps}C) reveals a
\textbf{no-free-lunch structure}: no configuration achieves high
sensitivity on both axes simultaneously --- the upper-right quadrant
is empty. Fingerprints cluster at zero stereo CV with high MMP CV;
MolFormer cosine sits at high stereo CV with intermediate MMP CV;
ChemBERTa is poor on both. The optimal representation depends on
which structural type of cliff is of interest.

\section{Discussion}
One of the aims of this experiment was to show, that the choice of the (embedding, metric) pair 
has a substantial effect on the activity cliff counts. The most striking effect is probably the 
change in activity cliff counts on a fixed dataset by up to 55-fold.
Hence, the representation effect is substantially larger than any biological 
effect observed across the three protein targets in our benchmark.
Indeed, the same dataset contains 142 cliff pairs
at 90\% Morgan Tanimoto similarity, 752 under Chemeleon cosine, and
7\,903 under RDKit Dice. These are three valid answers to three
different questions about molecular proximity. Published comparisons of
cliff prevalence that do not control for representation compare
convolutions of chemistry and geometry. The pipeline introduced here
aims in quantifying the effect of embedding space geometry.

\subsection*{Different representations encode different structural
scales}

Classical fingerprints and learned embeddings are not competing
approximations to the same ideal --- they encode non-overlapping
aspects of molecular structure. \textbf{Morgan fingerprints} organizes
space around scaffold identity: bimodal Beta-distributed distances,
the strongest cliff enrichment ($G = 0.077$),
but complete stereochemical blindness. \textbf{MACCS and RDKit Dice}
are sensitive at the transformation scale (MMP CV $0.69$--$0.99$,
the highest in the benchmark), encoding functional-group patterns
at exactly the fragment scale at which single-step MMP transformations
manifest --- but carry the same stereochemical blindness and, for
RDKit, topological collapse on Factor Xa. \textbf{MolFormer cosine}
encodes stereocentre information as directional variation in embedding
space: stereoisomer CV $1.18$--$1.22$ (the most stable across all
targets), the only configuration that can detect enantiomeric cliffs.
\textbf{Chemeleon} produces the richest topological cliff structure
(Steps~3--4), but with a
dramatic metric-dependence: Chemeleon $L^1$/$L^2$ collapse
topologically on Dopamine D2 while Chemeleon cosine retains the
highest mean persistence on that target --- metric choice alters
global geometry without affecting locally discriminative structure.

Across both stereo and MMP analyses and both MolFormer and Chemeleon,
cosine distance consistently outperforms $L^1$/$L^2$. The mechanism
is coherent: stereochemical and transformation-relevant features appear
to be encoded as \textit{directional} variation in embedding vectors.
Cosine distance is sensitive to angular differences while invariant to
vector magnitude; Euclidean and Manhattan distances probably dilute this signal.
Cosine should be the default metric for learned embeddings in any
analysis sensitive to specific structural features.

\subsection*{Pipeline non-redundancy and extracting biological information}

Each step reveals failure modes invisible to the others: Step~2 exposes
RDKit's poor cliff placement despite its leading CV; Step~3 exposes its
heavy gradient tails despite acceptable enrichment; Step~4 reveals its
topological collapse on Factor Xa and a new failure of Chemeleon
$L^1$/$L^2$ on Dopamine D2. 

On the other hand, the convergence of Lorenz enrichment, KWW
tail severity, persistent homology, and predictive AUC on the same
target ordering (SARS-CoV-2 smoothest, Dopamine D2 roughest) is the
strongest evidence that this ordering reflects biology, not any single
diagnostic. It is a useful set of tools to distinguish the biological effect, from
the representation artifact. It is also interesting to speculate, why the Dopamine D2 landscape is rougher,
meaning the similar compounds are more likely to form a cliff. 
Dopamine D2 is a G protein-coupled receptor (GPCR). GPCRs are inherently flexible membrane proteins 
that exist in multiple conformational states —- inactive, active, and intermediate —- in dynamic equilibrium. 
A ligand's potency depends not just on its interactions with a single rigid binding site but 
on how it shifts this conformational equilibrium. Small structural changes can shift a ligand 
from stabilizing the inactive state to stabilizing the active state, producing enormous potency 
changes that have no simple relationship to structural distance in any fingerprint or embedding space.
Alternatively, Dopamine D2 landscape may be rougher, because D2 receptor accommodates ligands that 
simultaneously engage the orthosteric binding site (the catecholamine pocket) and the secondary 
binding pocket or extracellular vestibule. Yet another possibility is, that the reason lies in
the dataset being drawn from ChEMBL, aggregating measurements from many laboratories, assay conditions, 
and time periods. All of these factors are absent in the case of SARS-CoV-2 dataset, which has the 
most diverse set of ligands, not modifications of known scaffolds. It is reassuring to know,
that the biology of the target may be visible in the pure mathematical analysis of the data.

\subsection*{Practical recommendations}
The whole analysis shows, that for different tasks different representations are optimal. The proposed 
scenarios are gather in Table~\ref{tab:recommendations}.

\begin{table}[h]
\centering
\caption{Recommended (embedding, metric) configurations from the
present benchmark, by downstream task. All recommendations are
conditional on the target not requiring stereochemical discrimination
(for which MolFormer, or Chemeleon cosine is required regardless of task) and on
ChemBERTa-77M-MLM being excluded from consideration.}
\label{tab:recommendations}
\small
\begin{tabular}{L{3.8cm} L{3.2cm} L{6.8cm}}
\toprule
Task & Recommended configuration & Rationale \\
\midrule
Cliff identification and SAR analysis within a structural series &
  Morgan Tanimoto &
  Highest $G$ across all targets; \\[6pt]
Stereochemistry-sensitive programs (chiral candidates, enantiomeric
pairs) &
  MolFormer cosine &
  Only configuration with $p_{0,\text{stereo}} = 0$ and CV
  $> 1$ for stereoisomers; most stable cross-target (range $0.037$) \\[6pt]
MMP transformation annotation and single-step SAR &
  MACCS Dice or RDKit Dice &
  Highest MMP CV ($0.73$--$0.99$); encodes fragment-level
  transformations at the scale at which MMP changes manifest \\[6pt]
Global landscape exploration and topological analysis &
  Chemeleon cosine &
  Richest topological cliff structure on two of three targets;
  most stable against collapse; validate on a representative
  subset before full-scale use \\[6pt]
Assessing intrinsic landscape roughness of a new target &
  Run Steps~1--3 on all configurations &
  Universal $b < 2$ across all configurations is the
  representation-independent signature of intrinsic roughness;
  requires consensus across representations \\
\bottomrule
\end{tabular}
\end{table}

Another question is the threshold specifying which compounds are considered to be similar.
It seems that the field has converged on $\geq 0.85$ Tanimoto similarity with Morgan 
fingerprints.\cite{stumpfe2012exploring,maggiora2006outliers}
This convention is implicitly calibrated to Morgan fingerprint geometry and has no principled 
justification when applied to other representations. Here we propose three independent,
data-driven estimates of the representation-specific threshold range,
each derived from a different geometric property of the activity
landscape.

The first is the sigmoid inflection point $x_0$ of the instantaneous cliff rate
(Step~2, see Supplementary Information Section S2.4 for details) 
identifies the similarity percentile
at which cliff frequency transitions most rapidly toward the null
baseline. Below $x_0$, structural proximity reliably implies activity
consistency; above it, cliff frequency approaches the random baseline.
The second estimate is the parameter $p_\text{max}$ from the $H_0$ persistence diagram
(Step~4) which identifies the structural distance at which the most isolated
cliff cluster merges with the broader cliff population --- the largest
distance gap between a coherent cliff family and the surrounding
subspace. Setting $s \approx 1 - p_\text{max}$ captures the most
structurally interpretable cliff population while partially excluding
unstructured background pairs. The third estimate is the 95th-percentile gradient $r_{95}$
(Step~3) which constrains the ratio $A/s$ from below: a pair with
$\mathcal{L}(i,j) = A/s$ sits exactly at the cliff boundary, so
$s > A/r_{95}$ must hold for the threshold to capture any of the
most extreme gradient pairs.

\subsection*{Implications and future work}
The current work may be expanded in several ways. First, we fix the parameters of the representation,
and the splitting methodology. Varying Morgan radius or clustering-based split criteria would 
introduce additional degrees of freedom whose interaction with the five geometric diagnostics 
would require a dedicated study; we therefore fix these parameters at their conventional values to 
isolate the effect of embedding and metric choice, which is the variable of primary interest. Similarly,
one might include additional datasets, representations, or sample different activity thresholds, or 
optimize the model architecture. Each of these extensions would be for sure an interesting 
further research direction, yet it would not change the main claim of the current work - the 
observed quantities are representation-dependent, and the choice of the (embedding, metric) pair 
is crucial.

Maybe one interesting, genuinely new future direction follows from the observation, 
that there is no free lunch -- no configuration excells in all steps simultaneously.
This is partially by design of the representations -- the fingerprints and learned 
embeddings concentrate on different aspects of molecule structures. This suggests, that 
instead of searching for one ultimate representation one might combine them. For example, 
one might construct a dynamically weighted SALI index following 
from different representations, which might be an ultimate measure of the activity cliff landscape.

\bibliography{biblio}

\end{document}